\documentstyle [12pt]{article}
\begin{document}

\title{$p$--Adic Models of Ultrametric Diffusion Constrained by Hierarchical Energy Landscapes}
\author{V A Avetisov, A H Bikulov, S V Kozyrev, V A Osipov}
\date{}
\maketitle
\begin{quote}{\footnotesize {\it{N N Semenov}} Institute of Chemical Physics, Russian Academy of Sciences,
 Moscow, Russia}
\end{quote}

\begin{abstract}
We  demonstrate that $p$--adic analysis is a natural basis for
the construction of a wide variety of the ultrametric diffusion
models constrained by hierarchical energy landscapes. A general
analytical description in terms of $p$-adic analysis is given for
a class of models. Two exactly solvable examples, i.e. the
ultrametric diffusion constraned by the linear energy landscape
and the ultrametric diffusion with reaction sink, are considered.
We show that such models can be applied to both the relaxation in
complex systems and the rate processes coupled to rearrangenment
of the complex surrounding.
\end{abstract}

\section{Introduction}

The concept of hierarchical energy landscape attracts a lot of
interest in connection with  relaxation phenomena  in complex systems, in
particular, glasses, clusters, and proteins
\cite{Mezard1}-\cite{Frauenfelder3}. This concept can be
outlined as follows. A complex system is assumed to have a large
number of metastable configurations which realize local minima on
the potential energy  surface. The local minima
are clustered in hierarchically nested basins of minima, namely,
each large basin consists of smaller basins, each of these consisting
of even smaller ones, and so on.  To be more definite,
 the  hierarchy of  basins possesses  ultrametric geometry. Finally, the basins
of minima are separated from one another
 by a  hierarchically arranged set of
barriers, i.e.,  high barriers separate large basins, and smaller
basins within each larger one are separated by lower barriers.

Transitions between the basins are determined by rearrangements of
the system configuration for different time  scales.

Thus, two key points of the concept of hierarchical energy
landscape should be  marked. First, the configurational
 space of the system is approximated by an ultrametric space;
 second, the configurational
 rearrangements of the system are described
by  stochastic motion in the ultrametric configurational
 space.

This concept has been implemented in a number of toy models
referred to as random walks in   ultrametric space,  diffusion in
ultrametric space, and ultra-diffusion \cite{Ogielski1} -\cite{
Kohler1}. A recent term for such processes is ``basin-to-basin
kinetics'' \cite{Becker1}. In fact, all these models deal with a
certain type  of stochastic motion, which in this paper  we call
diffusion in ultrametric space, or shortly, ultrametric diffusion.

The toy models have definitely  demonstrated that ultrametricity
actually reflects the characteristic  features of relaxation in
complex systems. It is commonly recognized that relevant
analytical tools should be developed for applying the concept
of   hierarchical energy landscape to the description of
relaxation phenomena in complex systems \cite{Frauenfelder2} --
\cite{Becker1}, \cite{Frauenfelder3}. In this paper, we show that
$p$--adic analysis is an adequate analytical tool for studying
these problems.

The field of $p$--adic numbers is  the most important example of
ultrametric spaces  (see, for instance, \cite{Rammal}). An
introduction  to $p$--adic analysis can be found in
\cite{Vladimirov1}. The $p$--adic mathematical physics attracts a
great deal of interest in quantum mechanics, string theory,
quantum gravity \cite{Vladimirov1} -- \cite{Arefeva}, spin--glass
theory \cite{ABK, Parisi1}, and theoretical biology
\cite{Dubischar}. In works \cite{ABK, Parisi1} the $p$--adic
parameterization of the replica Parisi matrix was developed. In
papers \cite{Carlucci, Carlucci1} the $p$--adic Fourier
transformation (called there the replica Fourier transformation)
was applied in the replica method. In the work \cite{wavelet} it
was shown that the $p$--adic change of variables maps the basis of
eigenvectors of the Vladimirov operator of $p$--adic fractional
derivation onto the basis of wavelets in $L^2({\bf R})$.

We demonstrate here that $p$--adic analysis is a natural basis
for the construction of a wide variety of sufficiently complex
models, which can be efficiently used in applications to both the
relaxation in complex systems and the rate processes coupled to
the relaxation of complex environment.

This paper has the following structure. In section 2, a  general
analytical description  in terms of $p$--adic analysis is given
for  a class of ultrametric diffusion models, the so-called pure
models (see \cite{Rammal}). Since $p$--adic analysis seems to be a
fairly new  analytical tool, we thought it appropriate to
describe  the  technique of solving the corresponding $p$--adic
master equation. Specific examples are considered in the next two
sections. In section 3, we give a complete description of the
ultrametric diffusion constrained by the linear energy landscape.
 In section 4, we consider a special case  when
the rate process coupled to  the configuration rearrangements
can be described;
and we also introduce and study  a model of ultrametric
diffusion with a reaction sink.
We show that  such
models can be applied, in particular, to the ligand
rebinding kinetics in heme proteins.

\section{Ultrametric diffusion}

In this Section,  we describe a class of pure ultrametric
diffusion models corresponding  to regular hierarchical
landscapes with degeneracy. First, we review a  model of
diffusion on an ultrametric lattice \cite{Ogielski1} (this model
was considered in \cite{ABK} with the help of $p$--adic analysis).
Then, passing to a continuous description, we obtain the
$p$--adic master equation. The last part of this section contains
a brief description of the standard analytical technique for the
investigation of such $p$--adic equations.

\subsubsection*{Ultrametric lattice and the master equation}

Consider a system with the state space $B_N$ consisting of the
points $i = 1,\ldots,p^N$, where $p$ is a prime number. These
points can be regarded  as lattice sites. On this lattice, a
regular  hierarchical energy landscape can be constructed  as
follows (see figure 1). Let us divide the space $B_N$ into $p$
mutually disjoint  subsets (basins) $B_{N-1}(a_1)$, $a_1 =
1,\ldots,p$, where each $B_{N-1}(a_1)$ consists of  $p^{N-1}$
points: $\bigcup\limits_{a_1}B_{N-1}(a_1)=B_N\,.$

Let us introduce activation barriers. The activation barrier for
the transitions between the basins $B_{N-1}(a_1)$ with different
values of $a_1$ is taken equal to $H_N$. Thus, the probability of
transition between any two sites $i\in B_{N-1}(a_1)$, $j\in
B_{N-1}(a_1')$ of different basins $B_{N-1}$ is equal to
$\rho_N$. In the Arrhenius case, we have $\rho_N = e^{-{H_N\over
kT}}$. Further, let us divide each  basin $B_{N-1}(a_1)$ into $p$
smaller basins $B_{N-2}(a_1 a_2)$, $a_2 = 1,\ldots,p$, each
having $p^{N-2}$ sites: $\bigcup_{a_2} B_{N-2}(a_1
a_2)=B_{N-1}(a_1)$.

Assume that the activation barriers within each larger basin
$B_{N-1}(a_1)$  are equal to $H_{N-1}$. Moreover, assume that the
probability of transition between any two sites $i\in B_{N-2}(a_1
a_2)$, $j\in B_{N-2}(a_1'a_2')$  of different basins $B_{N-2}$ is
equal to $\rho_{N-1}$ for $a_1=a_1'$, and is equal to $\rho_N$
for $a_1\ne a_1'$. In other words, the probability of transition
between sites of different basins $B_{N-2}$ depends on the
hierarchical level, $N-1$ or $N$, at which these basins merge
into a single super basin. We proceed like this until we reach
the first level with each basin $B_1(a_1 a_2\ldots a_N)$ having
only one site. As a result, the probability of transition between
any two sites is  specified by a single value $\rho_{\gamma}$
among $\rho_1>\ldots>\rho_\gamma >\ldots >\rho_N$,  according to
the level $\gamma = 1,2,\ldots ,N$ on  which these two sites
happen to be in a single basin $B_\gamma$.

Denote by  $f_i(t)$ the probability of finding the system in the
site $i$ at time $t$. Using the equivalence between the
transition probability matrix and the Parisi matrix \cite{ABK,
Parisi1}, we can write the master equation for the evolution of
probabilities in the form \cite{ABK}
\begin{equation}\label{1}
\frac{d}{dt}{\bf f}(t)=\left( {\bf Q} - \lambda_0 {\bf I} \right)
{\bf f} (t),
\end{equation}
where ${\bf f}(t)=\left\{ f_1 (t),\ldots ,f_{p^N}(t) \right\}$ is
the state vector, ${\bf Q}$  is the transition $p^N\times p^N$
matrix of Parisi type \cite{Parisi2}, $\bf I$  is the unity
matrix, and $\lambda_0 =
\sum_{\gamma=1}^{N}p^{\gamma-1}\rho_\gamma$  is the eigenvalue of
the matrix ${\bf Q}$ corresponding  to the eigenvector with equal
components. For instance, the transition matrix ${\bf Q}$ for
$p=2$ has the form
\begin{equation}\label{matrix}
{\bf Q}=\left(
\begin{array}{ccccccccc}
{0}&\rho_{1}&\rho_{2}&\rho_{2}&\rho_{3}&\rho_{3}&\rho_{3}&\rho_{3}&\dots\\
\rho_{1}&{0}&\rho_{2}&\rho_{2}&\rho_{3}&\rho_{3}&\rho_{3}&\rho_{3}&\dots\\
\rho_{2}&\rho_{2}&{0}&\rho_{1}&\rho_{3}&\rho_{3}&\rho_{3}&\rho_{3}&\dots\\
\rho_{2}&\rho_{2}&\rho_{1}&{0}&\rho_{3}&\rho_{3}&\rho_{3}&\rho_{3}&\dots\\
\rho_{3}&\rho_{3}&\rho_{3}&\rho_{3}&{0}&\rho_{1}&\rho_{2}&\rho_{2}&\dots\\
\rho_{3}&\rho_{3}&\rho_{3}&\rho_{3}&\rho_{1}&{0}&\rho_{2}&\rho_{2}&\dots\\
\rho_{3}&\rho_{3}&\rho_{3}&\rho_{3}&\rho_{2}&\rho_{2}&{0}&\rho_{1}&\dots\\
\rho_{3}&\rho_{3}&\rho_{3}&\rho_{3}&\rho_{2}&\rho_{2}&\rho_{1}&{0}&\dots\\
\vdots&\vdots&\vdots&\vdots&\vdots&\vdots&\vdots&\vdots&\ddots\\
\end{array}
\right).
\end{equation}
This form of the transition probability matrix ${\bf Q}$ is a
direct consequence of the hierarchical picture of
basin--to--basin transitions.

The  master equation (\ref{1}) coincides with the equation
of ultrametric diffusion considered in \cite{Ogielski1}.

\subsubsection*{Continuum limit}

The continuum limit for the  above hierarchical lattice model is
discussed in \cite{ABK} and \cite{Parisi1}. The results obtained
in \cite{ABK} can be summarized as follows. We introduce the
following  one-to-one mapping of integers $ i$  enumerating the
lattice sites onto a set of real numbers $x$:
$$
i=1+p^{-1}\sum_{j=1}^n x_j^{(i)} p^j\to\sum_{j=1}^N
x_j^{(i)}p^{-j}=x^{(i)},\qquad 0\leq x_j^{(i)}\leq p-1.
$$

The value $x^{(i)}$ is regarded as a coordinate of the site $i$.
Then, in the continuum limit, the set of coordinates $x$ is
extended to the field of $p$--adic numbers $Q_p$, the matrix
$\left( {\bf Q} - \lambda_0 {\bf I} \right)$ turns into an
integral operator, and the master equation (\ref{1}) transforms
into the following equation:
\begin{equation}\label{3}
\frac{d}{dt}f(x,t)=\int_{Q_p}(f(y,t)-f(x,t))\rho(|x-y|_p)
d\mu(y)\;.
\end{equation}
Here, the operator on the right--hand side is called the
ultrametric diffusion operator, the function $\rho(|x-y|_p)$:
$Q_p \times Q_p \mapsto {\bf R}$ (${\bf R}$ is the field of real
numbers) is the kernel of the ultrametric diffusion operator, the
function $f(x)$: $Q_p \mapsto {\bf R}$ is  the probability
density distribution, and $d\mu(y)$ is the Haar measure on the
field of $p$--adic numbers~$Q_p$.

Expression (\ref{3}) specifies the general form of ultrametric
diffusion operators. A~special case of the operator on the
right-hand side of (\ref{3}) should be mentioned, namely, the
Vladimirov operator
\begin{equation}\label{4}
D^{\alpha}_x f(x,t)=\frac{1}{\Gamma_p(-\alpha)}
\int_{Q_p}\frac{f(x)-f(y)}{|x-y|_p^{1+\alpha}}d\mu(y)\;,
\end{equation}
where $\alpha>0$ and
$\Gamma_p(\alpha)=\displaystyle{\frac{1-p^{\alpha-1}}{1-p^{-\alpha}}}$
is the $p$--adic gamma function. This operator is an analog of the
differentiation operator in $p$--adic analysis. For this reason,
the equation
\begin{equation}\label{5}
\frac{d}{dt}f(x,t)=-D^{\alpha}_x f(x,t)
\end{equation}
is interpreted in the $p$--adic mathematical physics as the
equation of Brownian motion on a $p$--adic line. More details
concerning pseudo--differential operators and equation (\ref{5})
can be found in the monograph \cite{Vladimirov1}. It is
interesting to note that an  equation of type (\ref{5}) is the
master equation for the model in which the height of the barriers
separating the basins of states has linear growth with respect to
the hierarchy level number \cite{ABK}.

\subsubsection*{Solution Technique}

Next,  we give a brief review of  the standard technique for
solving equations of type (\ref{3}). We investigate the Cauchy
problem for  equation (\ref{3}) with the initial condition
$f(x,0)=\delta(x)$, where $\delta(x)$ is the $p$--adic delta
function \cite{Vladimirov1}. Let us find a fundamental solution of
equation (\ref{3}). Applying the $p$--adic Fourier transformation
to (\ref{3}), we obtain the following equation:
\begin{equation}\label{6}
\frac{\partial \tilde f (\xi,t)}{\partial t}=-\tilde \rho
(|\xi|_p) \tilde f(\xi,t)\;,
\end{equation}
where
\begin{equation}\label{7}
\tilde \rho (|\xi|_p)=\int_{Q_p} \rho(|x-y|_p) (1-\chi(\xi x))
d\mu(x)\;,
\end{equation}
and $\tilde f(\xi,t)$  is the Fourier transform of the
distribution $f(x,t)$. Hence, it is easy to find  the Fourier
transform
\begin{equation}\label{8}
\tilde f (\xi,t)=\exp(-\tilde\rho(|\xi|_p)t)\;.
\end{equation}
Applying the inverse Fourier transformation to (\ref{8}), we
obtain the fundamental solution
\begin{equation}\label{9}
f(x,t)=\int_{Q_p}\exp(-\tilde\rho(|\xi|_p)t) \chi(-\xi x)
d\mu(\xi)\;,
\end{equation}
where $\chi(\xi x)$  is the additive character of the field of
$p$--adic numbers.

Equation (\ref{3}) is defined on the whole field of $p$--adic
numbers $Q_p$. Therefore, the solution (\ref{9}) describes the
diffusion process in an unbounded ultrametric space. However, for
some applications, it is important to study the problem of
ultrametric diffusion in a  bounded region of the state space. In
this case, the following procedure can be used.

Consider  diffusion in the $p$--adic disc $B_r=\{x:|x|_p\le
p^r\}$  of radius  $p^r$ with center at zero. In this case, the
master equation reads
\begin{equation}\label{10}
\frac{\partial f (x,t)}{\partial t}=
\int_{B_r}(f(y,t)-f(x,t))\rho(|x-y|_p) d\mu(y)\;.
\end{equation}
Changing the variables, $x=p^{-r}z$, we transform this equation to
$$
\frac{\partial  f (z,t)}{\partial t}=
\int_{Z_p}(f(z',t)-f(z,t))\rho(|z-z'|_p) d\mu(z')\;.
$$

This equation can be investigated by means of $p$--adic Fourier
series
$$
\phi(z,t)=\sum_{k\in I}\tilde\phi_k(t)\chi(-kz)\;,
$$
where the characters $\{\chi(kz)\}$ form an orthonormal basis in
$L^2(Z_p)$ labeled by
$$
k=0,\quad k=p^{-\gamma}\left(
k_0+k_1p+\ldots+k_{\gamma-1}p^{\gamma-1} \right)
$$
$$
\gamma=1,2,\ldots;\quad k_0=1,2,\ldots,p-1;\quad
k_j=0,1,\ldots,p-1;\quad j=1,2,\ldots,\gamma-1\;.
$$

\section{Diffusion on linear hierarchical landscape}
In this section, we examine  ultrametric diffusion in the case of
a linear landscape  with  the height of activation barriers
$H_\gamma$ having linear growth with respect to the number
$\gamma$ of the hierarchical level. Since $|x-y|_p=p^\gamma$, for
the linear landscape we have
\begin{equation}\label{11}
H_\gamma=H_0 \ln(|x-y|_p)\;,
\end{equation}
where $H_0$ is a scale parameter. Suppose that the probability of
transition between the states separated by the activation barrier
$H_\gamma$ is defined by
\begin{equation}\label{12}
\rho_\gamma=w_0 p^{-\gamma}  \exp\left(-\frac{H_\gamma}
{kT}\right)\;,
\end{equation}
where $w_0$ is a pre-exponential factor, $T$ is the temperature,
$k$ is the Boltzmann constant. As noted in the previous section,
under these assumptions, the kernel of the ultrametric diffusion
operator coincides with the kernel of the Vladimirov operator
(see also \cite{ABK}):
$$
\rho(|x-y|_p)=\frac{w_0}{|x-y|_p^{\alpha+1}} \;,
$$
where
$\alpha=H_0/kT$. The Fourier transform of this kernel is
\begin{equation}\label{13}
\tilde\rho(\xi)=-w_0 \Gamma_p(-\alpha)|\xi|_p^{\alpha}.
\end{equation}

Thus, in the case of a linear hierarchical landscape (\ref{11})
(under the assumption (\ref{12})), the master equation of
ultrametric diffusion coincides with the equation of  Brownian
motion on the $p$-adic line \cite{Vladimirov1} (see relations
(\ref{4}) and (\ref{5}) in section~2):
\begin{equation}\label{14}
\frac{\partial f(x,t)}{\partial t}-w_0
\Gamma_p(-\alpha)D^{\alpha}_x f(x,t)=0\;.
\end{equation}

Let us examine the solutions of this equation. As the the initial
value we take $f(x,0)$ which is constant inside a bounded set in
$Q_p$ and vanishes outside. For instance, we can take as the
initial function  the indicator  of $Z_p$:
\begin{equation}\label{15}
f(x,0)=\Omega(|x|_p)=\left\{
\begin{array}{rc}
1,&|x|_p\le 1\;,\\
0,&|x|_p>1\;.\\
\end{array}
\right.
\end{equation}
Note that the $p$--adic Fourier transform of this indicator
function is the same function. The diagram of the model with
this  initial condition is given in figure 2. Using the
fundamental solution (\ref{9}), one can construct the solution of
equation (\ref{14}) with the initial condition (\ref{15}):
\begin{equation}\label{16}
f(x,t)=\int_{Q_p}\chi(-\xi x)\Omega(|\xi|_p)\exp\{
\Gamma_p(-\alpha)|\xi|_p^{\alpha}w_0t\}\;d\mu(\xi).
\end{equation}
Calculating the integral, we get
$$
f(x,t)=(1-p^{-1})|x|_p^{-1}\sum_{\gamma=0}^{+\infty}
p^{-\gamma}\Omega\left(\frac{p^{-\gamma}}{|x|_p}\right)
\exp\left\{ \frac{p^{-\alpha\gamma}}{|x|_p^\alpha}\Gamma_p(-\alpha)w_0t\right\}-\\
$$
\begin{equation}\label{17}
\qquad{ } -|x|_p^{-1}\Omega\left(\frac{p}{|x|_p}\right)
\exp\left\{
\frac{p^\alpha}{|x|_p^\alpha}\Gamma_p(-\alpha)w_0t\right\}\;.\\
\end{equation}

The solution (\ref{17}) describes  the  diffusion process in the
unbounded ultrametric space. As an illustration, the
time-dependence  of population densities for different states is
shown in figure 3. It is easy to see that at fairly low
temperatures (with $\alpha$ much larger than $1$),  the decay of
the population density of the states has a discrete character due
to the hierarchy of the characteristic times of transition
through the respective activation barriers. As the temperature
increases, the discrete  nature of decay becomes less and less
noticeable and at  sufficiently high temperatures (with $\alpha$
of the order of or less than 1) the population density of the
states is governed by the power law.

This conclusion can also be  obtained  by analytical means. Now,
let  us estimate  the solution (\ref{17}) for $|x|_p>1$. Using
the Abel transformations
$$
f(x,t)=|x|_p^{-1}\sum_{\gamma=0}^{+\infty}p^{-\gamma}\left[
\exp\left\{
\frac{p^{-\alpha\gamma}}{|x|_p^\alpha}\Gamma_p(-\alpha)w_0t\right\}-
\exp\left\{\frac{p^{-\alpha(\gamma-1)}}{|x|_p^\alpha}\Gamma_p(-\alpha)w_0t\right\}\right]=
$$ $$ \qquad{ } =|x|_p^{-1} \sum_{\gamma=0}^{+\infty}p^{-\gamma}
\int_{\gamma-1}^{\gamma}{d\over dz}\left[ \exp\left\{ p^{-\alpha
z}|x|_p^{-\alpha}\Gamma_p(-\alpha)w_0t\right\} \right]dz $$ and
the estimate $$ p^{-m}\int_{m-1}^{m}g(z)dz\le
\int_{m-1}^{m}p^{-z}g(z)dz\le p^{-m+1}\int_{m-1}^{m}g(z)dz\;, $$
we get $$
\frac{(w_0t)^{-1/\alpha}}{p(-\Gamma_p(-\alpha))^{1/\alpha}}
\gamma\left(1+\frac{1}{\alpha}, -\Gamma_p(-\alpha)w_0t
\left(\frac{p}{|x|_p}\right)^{\alpha}\right)\le f(x,t)\le $$ $$
\qquad{}\le\frac{(w_0t)^{-1/\alpha}}{(-\Gamma_p(-\alpha))^{1/\alpha}}
\gamma\left(1+\frac{1}{\alpha}, -\Gamma_p(-\alpha) w_0 t
\left(\frac{p}{|x|_p}\right)^{\alpha}\right)\;,\\
$$ where $\gamma(a,b)$ is the incomplete gamma function. If
$$
w_0t\gg
\left(\frac{|x|_p}{p}\right)^{\alpha}\frac{1}{-\Gamma_p(-\alpha)},
$$ then $$ \gamma\left(1+\frac{1}{\alpha},-\Gamma_p(-\alpha)w_0t
\left(\frac{p}{|x|_p}\right)^{\alpha}\right)\approx\Gamma\left(1+\frac{1}{\alpha}\right),
$$ where $\Gamma(a)$ is the gamma function. Thus, for any state
$x$, there is a characteristic time $$
\tau(x,\alpha)=\left(\frac{|x|_p}{p}\right)^{\alpha}\frac{w_0^{-1}}{-\Gamma_p(-\alpha)}
$$ such that  the population density $f(x,t)$, for $t\gg
\tau(x,\alpha)$, can be estimated by power functions:
\begin{equation}\label{18}
\frac{\Gamma\left(1+\frac{1}{\alpha}\right)}
{p(-\Gamma_p(-\alpha))^{1/\alpha}}(w_0t)^{-1/\alpha}\le
f(x,t)\le
\frac{\Gamma\left(1+\frac{1}{\alpha}\right)}
{(-\Gamma_p(-\alpha))^{1/\alpha}}(w_0t)^{-1/\alpha}
\end{equation}

Let us calculate the average of  the distribution (\ref{17}). Note
that the moment integrals
\begin{equation}\label{19}
\langle |x|_p^\beta\rangle=\int_{Q_p}|x|_p^\beta f(x,t)d\mu(x)
\end{equation}
are divergent  for all  $\beta\ge\alpha$,  and  therefore,
neither the mean $p$--adic distance nor the mean square of the
$p$--adic distance contain any information on the distribution
$f(x,t)$ if, for instance, $\alpha\le 1$. To describe the
ultrametric diffusion process in terms of mean values, some
meaningful characteristics should be introduced. The mean value
of the activation barrier to be overcome by the system by the
instant $t$ can be taken as one of such characteristics. In the
case of a linear hierarchical landscape (see (\ref{11})), the
mean value of the activation barrier is
\begin{equation}\label{20}
\langle H(t)\rangle=H_0\int_{Q_p}\ln |x|_p f(x,t)d\mu(x).
\end{equation}

Note that the integral on the right side of (\ref{20}) should be
understood as the limit
$$
\int_{Q_p}\ln |x|_p f(x,t)d\mu(x)=
\lim_{n\to +\infty}\int_{|x|_p\le p^n}\ln |x|_p f(x,t)d\mu(x) \;.
$$
Let
\begin{equation}\label{21}
L^{(n)}(t)=\int_{|x|_p\le p^n}\ln |x|_p f(x,t)d\mu(x) \;.
\end{equation}
By (\ref{16}) we get
$$
L^{(n)}(t)=p^n(n(1-p^{-1})-p^{-1}) \ln p
\sum_{\gamma = -\infty }^{-n} p^\gamma \exp\{
\Gamma_p(-\alpha)p^{\alpha\gamma}w_0t\}-
$$
$$
\qquad{} - \ln p
\sum_{\gamma =-n+1  }^{+\infty} \exp\{
\Gamma_p(-\alpha)p^{\alpha\gamma}w_0t\}\;.
$$
Therefore, the
quantities $L^{(n)}(t)$ satisfy the recurrent relations $$
L^{(n)}(p^{\alpha}t)=L^{(n-1)}(t)+p^{n-1}(1-p^{-1})\ln p
\sum_{\gamma = -\infty }^{-n+1} p^\gamma \exp\{
\Gamma_p(-\alpha)p^{\alpha\gamma}w_0t\}\;.
$$
Passing to the
limit as $n\to +\infty$, we obtain the following functional
equation:
$$
L(p^\alpha t)=L(t)+\ln p
$$
with the solution
$$
L(t)=\alpha^{-1}\ln t \;.
$$
Therefore,
\begin{equation}\label{22}
\langle H(t) \rangle =k T H_0^{-1}\ln t.
\end{equation}

Thus, the linear hierarchical landscape has a  remarkable
feature, namely,  the mean value of the activation barrier to be overcome
by the instant  $t$  has logarithmic growth with respect to  time.

Summing up the results of this section,  we  emphasize three
important features of ultrametric diffusion. The first and
foremost is that the  evolution of the population density in any
state has a discrete  nature and is determined by the hierarchy
of   the  relaxation time scales. This feature stems from the
concept of the hierarchical energy landscape and  is typical for
ultrametric diffusion. The discrete pattern of ultrametric
diffusion can be observed at sufficiently low temperatures (with
the dependence of $\alpha$ on $T$ considered above). Next, in the
case of a linear hierarchical landscape,  the long-time behavior
of the population density of states is described by the power
law.  Finally, the mean value of activation barriers    overcome
by a point by the time $t$ is logarithmically increasing with the
time.

For glasses and proteins, relaxation processes with similar properties
are discussed in  \cite{Binder,Ansary},
\cite{Frauenfelder1} -- \cite{ Becker1}, \cite{Steinbach} -
\cite{Leeson3}.

\section {Rate processes coupled to  configurational
rearrangements}

In this  section,  we apply the $p$--adic analysis to describe
the kinetics of transformations coupled to  configurational
rearrangements of the  surrounding.  We consider the well-known
example of such processes, namely, the ligand rebinding kinetics
of the myoglobin (Mb)
\cite{Ansary,Frauenfelder1,Frauenfelder2,Steinbach,Nienhaus}. In
this case, the object of investigation is the rebinding process
governed by  conformational  rearrangements of the protein
itself.  These investigations were the origin of the concept of
hierarchical energy landscape in the conformational  dynamics of
proteins.

The experiment is as follows. The myoglobin is transformed by
the laser pulse from the ground state [Mb-CO] into the
excited unbound state [$\rm {Mb}^*$-]. Then the protein macromolecule relaxes
into the functionally active  state [$\rm{Mb}_1$-].
From this state, the myoglobin
can go back to the ground state [Mb-CO]  owing to the binding of  the CO.

This process can be represented as the following sequence
$$
\mbox{Mb-CO}\stackrel{h\nu}{\longrightarrow}[\rm{Mb}^*\to\ldots\to\rm{Mb}_1]
\stackrel{\rm{CO}}{\longrightarrow}\mbox{Mb-CO}
$$

The rebinding kinetics is described by the survival probability
function $S(t)$, i.e.,  the probability of the fact that myoglobin
is  still unbound by the time $t$  after being exposed to the
laser pulse. Thus, the   quantity observed in these experiments
is  the total population of the conformational states
[$\rm{Mb}^*\to\ldots\to\rm{Mb}_1$] through which the protein
macromolecule  relaxes  from the excited state [$\rm{Mb}^*$-] to
the functionally active state [$\rm{Mb}_1$-].

Similar-type processes correlate with the models of ultrametric diffusion
described by the master  equation in the following form:

\begin{equation}\label{24}
\frac{\partial f(x,t)}{\partial
t}=\int_{Q_p}(f(y,t)-f(x,t))\rho(|x-y|_p) d\mu(y)
+\int_{Q_p}K(x,y)f(y,t) d\mu(y)\;.
\end{equation}
The last integral on the right-hand side of (\ref{24}) can be
thought of as a reaction sink distributed over  a bounded region
in the conformational space.  In this case, the survival
probability function is
\begin{equation}\label{25}
S(t)=\int_{Q_p}f(x,t)d\mu(x)\;.
\end{equation}

The procedure of solving  equations of  type (\ref{24}) is as
follows. Let us take $f(x,0)=f_0(x)$ in the initial condition.
Consecutively applying the $p$--adic Fourier transformation in $x$
and the Laplace transformation in the real  variable $t$, from
(\ref{25}), we get
\begin{equation}\label{26}
\hat{\tilde {f}}(\xi,s)=\frac{\tilde f_0(\xi)}{s+\tilde\rho
(|\xi|_p)}+\frac{1}{s+\tilde\rho(|\xi|_p)}\tilde
K(\xi,\zeta)f(\zeta,s) d\mu(\zeta)\;.
\end{equation}
where the symbol $ \hat{ } $ marks  the Laplace transform with
respect to the variable $s$, and $\tilde\rho (|\xi|_p)$  is given
by (\ref{7}). Next, we note that equation (\ref{26}) is a
heterogeneous Fredholm equation of the second kind.

Consecutively applying the inverse Laplace  transformation  and
the inverse $p$--adic Fourier transformation  to $\hat{\tilde
f}(\xi,s)$, we obtain  the solution $f(x,t)$ and the function
$S(t)$.

Let us give an example of a  model of this type (see figure~4)
  by making the following assumptions:
\begin{enumerate}
\item
Conformational rearrangements are restricted  to a bounded
region  of the conformational space of the system and this region
is the $p$--adic disc $B_r=\{x:|x|_p\le p^r,r\gg1\}$;

\item
The hierarchical landscape is linear, i.e., conformational
rearrangements are described by the ultrametric diffusion
operator of  type (\ref{4}) defined on  the disc $B_r$;
\item
In the region  $B_r$, there is a certain set of conformational
states in which the system can undergo  irreversible
transformations. This set of states  we describe by the  unit
disc $Z_p$.  In other words, there is a reaction--type sink at
each point of the disc $Z_p$;
\item
The rate of the reaction sink at each point of the disc $Z_p$ is
proportional to the mean value of the population density on that
disc;
\item
The initial distribution  $f(x,0)$ is constant on the the
$p$--adic layer
$$
R _{\sigma\delta}=\{x:p^{\delta+1}\le |x|_p\le
p^\sigma,\,0\le\delta<\sigma<r\}
$$
and vanishes outside the layer.
\end{enumerate}

The master equation for the model with the above assumptions is
\begin{equation}\label{27}
\frac{\partial f(x,t)}{\partial
t}=\int_{B_r}\frac{f(y,t)-f(x,t)}{|x-y|_p^{\alpha+1}} d\mu(y)
-\lambda\Omega(|x|_p)\int_{B_r}\Omega(|y|_p)f(y,t) d\mu(y)\;
\end{equation}
and the initial consitions have the form $$
f(x,0)=AG_\delta^\sigma(x)\equiv A\left[\Omega(|x|_p
p^{-\sigma})-\Omega(|x|_p p^{-\delta}) \right], \quad
A^{-1}=p^\sigma -p^\delta,\quad 0\le\delta<\sigma, $$ where
$\lambda$ is the rate parameter of the reaction sink. Let us use
the technique of Section~2 to solve  equation (\ref{27}). First,
we  pass from  the disk $B_r$  to  the  disk $Z_p$  by letting
$x=p^{-r}z$. Then the Cauchy problem takes the form
$$
\frac{\partial \phi (z,t)}{\partial
t}=p^{-r\alpha}\int_{Z_p}\frac{\phi(z',t)-\phi(z,t)}
{|z-z'|_p^{\alpha+1}} d\mu(z')-\\
$$
$$
\qquad{ } -\lambda p^r\Omega(|z|_p p^r)\int_{Z_p}\Omega(|z'|_p p^r)\phi(z',t)d\mu(z')\\
$$
$$
\phi(z,0)=AG_{\delta-r}^{\sigma-r}(z)\;,
$$
Using the
$p$--adic Fourier series for $Z_p$ (see Section~2) and reasoning
as above, we  obtain a system of equations for the Laplace
transforms  $\hat{\tilde{\phi}}_k(s)$ of the coefficients  of the
$p$--adic Fourier series:
$$
k=0: \quad
s\hat{\tilde{\phi}}_0(s)=p^{-r}-\lambda p^{-r}\left[
\hat{\tilde{\phi}}_0(s)+\sum_{q\in I\setminus \{0\}}\Omega(|q|_p
p^{-r})\hat{\tilde{\phi}}_q(s) \right]\;, $$ $$ k\ne 0: \quad
s\hat{\tilde{\phi}}_0(s)=A\tilde
G_{\delta-r}^{\sigma-r}(k)+p^{-\alpha r}(g_\alpha-|k|_p^\alpha)-
$$
$$
\qquad{} -\Omega(|k|_p p^{-r})\lambda p^{-r}\left[
\hat{\tilde{\phi}}_0(s)+\sum_{q\in I\setminus \{0\}}\Omega(|q|_p
p^{-r})\hat{\tilde{\phi}}_q(s) \right]\;,
$$
where
$$
g_\alpha=\frac{1-p^{-1}}{1-p^{-(1+\alpha)}}\;,\quad \tilde
G_{\delta-r}^{\sigma-r}(k) =p^{\sigma-r}\Omega(|k|_p
p^{\sigma-r})-p^{\delta-r}\Omega(|k|_p p^{\delta-r})\; .
$$

It is easy to find the solution to this system and write  the
equations for $\hat{\tilde{\phi}}_0(s)$ and
$\hat{\tilde{\phi}}_k(s)$.  If we note that in this case the
following equations hold:
$$
S(t)=p^r
\int_{Z_p}\phi(z,t)d\mu(z)=p^r\tilde\phi_0(t)\;, $$ $$ \hat
S(s)=p^r\hat{\tilde{\phi}}_0(s)\;,
$$
we can find the Laplace
transform of the survival probability function,
$$
\hat S(s)=\frac{1}{s+\lambda p^{-r}}
\left[1+\lambda^2p^{-2r}\frac{J_r(s)}{s+\lambda p^{-r}(1+sJ_r(s))} \right]
-\\
$$
\begin{equation}\label{28}
\quad -\frac{\lambda
A\left[p^{\sigma-r}J_{r-\sigma}(s)-p^{\delta-r}J_{r-\delta}(s)\right]}
{s+\lambda p^{-r}(1+sJ_r(s))}\;,
\end{equation}
where
$$
J_l(s)=\sum_{k\in I\setminus \{0\}}\frac{\Omega(|k|_p
p^{-l})}{ s+p^{-\alpha r}(|k|^{\alpha}_p-g_\alpha)}\;.
$$
Applying the inverse Laplace transformation to (\ref{28}), we find
$S(t)$.

The numerical results are as follows. The typical curves $S(t)$
within a certain time window are shown in figure~5 and  6. It is
clearly seen that there are three characteristic kinetic modes.
At high temperatures (curve 1 in Fig 5), the kinetics $S(t)$ is
governed by the exponential law. This mode is observed in the
case of $\tau_{eq}\ll  \lambda^{-1}$,  where $ \tau_{eq}$  is the
characteristic time during  which the  distribution $f(x,t)$
reaches quasi-equilibrium, and $\lambda$ is the reaction sink
parameter. In this case, during the observation periods
$t\gg\tau_{eq}$, the function $f(x,t)$ is close to a homogeneous
distribution  and the population of the reaction sink area (disc
$Z_p$) is equal to $p^{-r}S(t)$. Therefore, $S(t)$ is determined
by the kinetic equation $$ \frac{dS(t)}{dt}=-\lambda p^{-r}S(t),
$$ which describes the exponential relaxation
$S(t)=\exp(-t/\tau)$ with the relaxation time $\tau=p^r/\lambda$.
Since the limiting kinetic stage here is the reaction sink, this
mode may be called the reaction control mode. As the temperature
decreases (curves 2 and 3 in figure~5), a small section of  power
relaxation appears on the curve $S(t)$ and then extends to the
major part of the time window of observation. This change in the
curves $S(t)$ corresponds to the transition regime for which,
with the decrease of temperature, the time of relaxation to
equilibrium, $\tau_{eq}$, and the time parameter of the reaction
sink  $\lambda^{-1}$ become commensurable. In this case, the
kinetics of $S(t)$  is limited  first by the ultrametric
diffusion (power section of the curve $S(t)$), and then by the
reaction sink (the exponential decay). Note that if the time
window under observation shows  only the power section of the
curves $S(t)$, it seems that the decay rate $S(t)$ is growing
with the decrease of temperature.

With  further decrease of temperature (curves 1-3 in figure~6), a
beak-shaped section appears on the curve $S(t)$ which then
extends to  the entire time window under  observation. This mode
corresponds to situations with $\tau_{eq}\gg\lambda^{-1}$. In
this case, the decay rate  $S(t)$ is limited by filling of the
reaction sink area due to ultrametric diffusion. It is natural
to  call such a  mode the diffusion control mode. Now, there are
no "anomalous" kinetic effects in the time window under
observation, the rate of decay $S(t)$ falls down as the
temperature decreases.

It is interesting to note that all peculiarities  of the kinetic
curves for the model considered above are also typical for  the
ligand rebinding kinetics of myoglobin \cite{Ansary, Steinbach}.
Naturally, this gives grounds for some optimism. Nevertheless, we
would like to stress that the main objective of this paper is  to
demonstrate  a method for the construction of ultrametric
diffusion models based on $p$--adic analysis. The examples given
above should be considered as a mere illustration of the
possibilities offered by this approach, rather than particular
models giving a quantitative description of  particular
experiments. The development of such models  seems to be a
special task in each particular case.

\vspace{10mm}
{\bf Acknowledgements.}
We express  our gratitude to Prof. S.F. Fischer and Prof. Yu.A. Berlin for
useful and stimulating discussions and the support given to our group
 during  the early  stage of this research at the Joint Theoretical Group in
Munich. We would like to thank Prof. V.S. Vladimirov and Prof. I.V. Volovich
for valuable comments.
This work has been partly  supported by INTAS ( grant No. 9900545),
and  The Russian Foundation for Basic Research (projects
990100866 and  001597392).

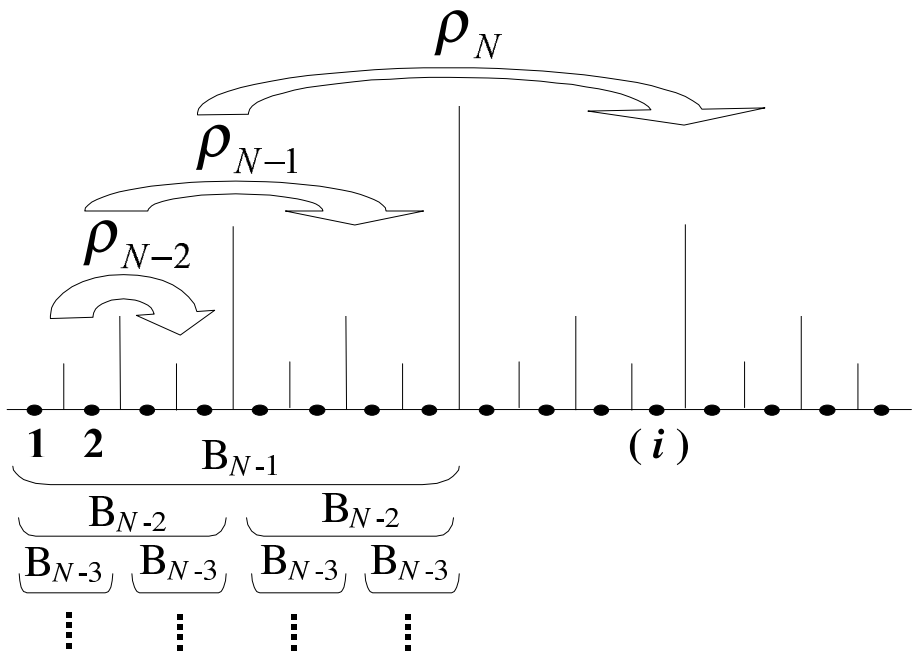
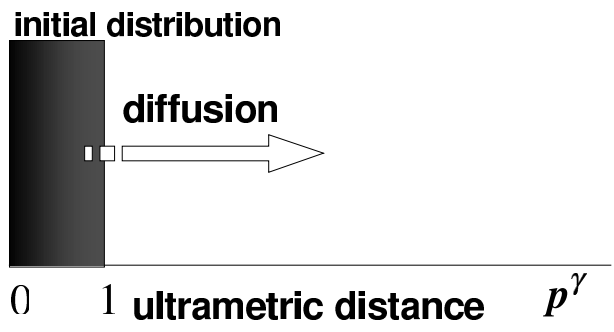
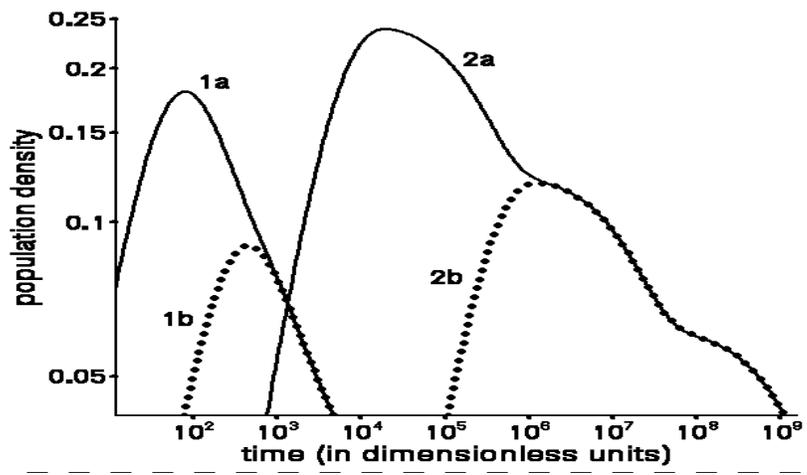
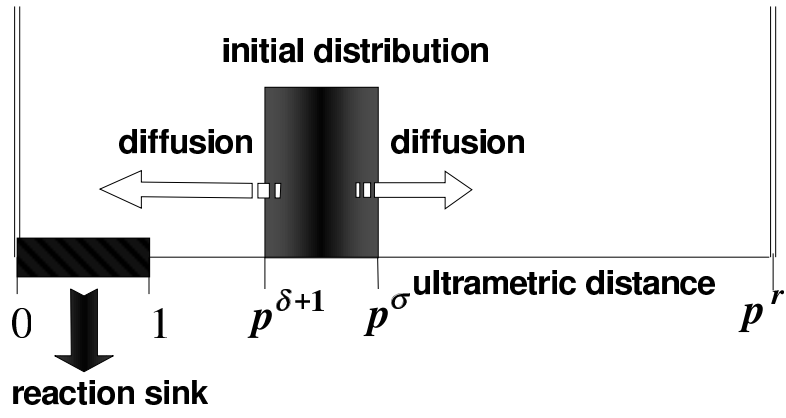
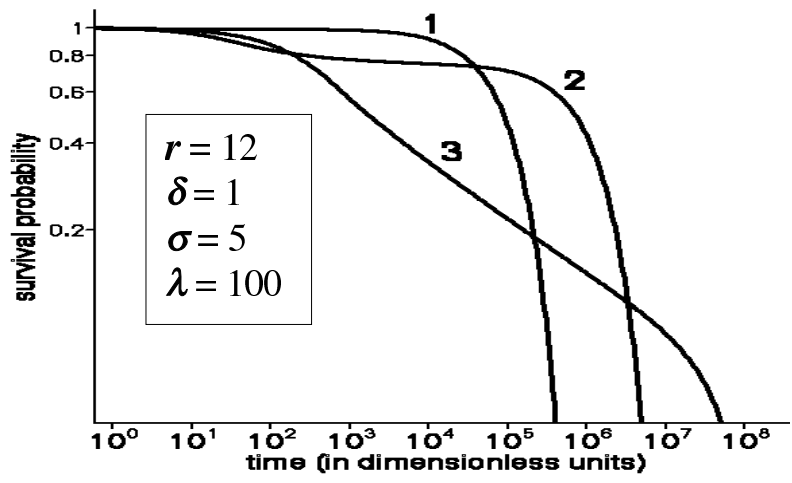
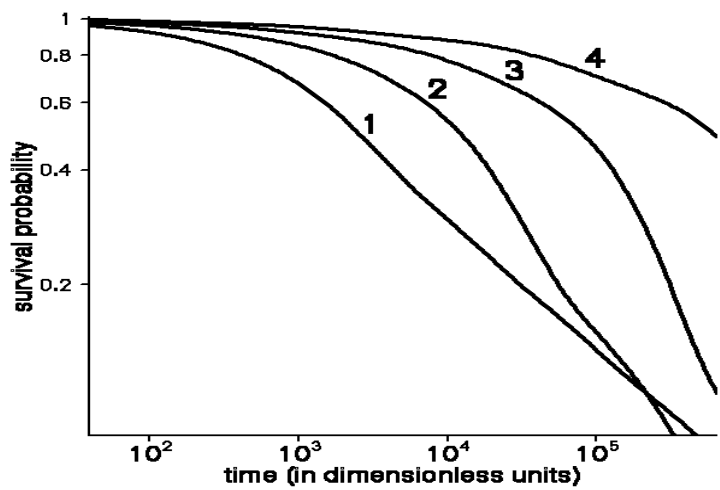
\end{document}